\documentclass[preprint,onecolumn,floatfix,superscriptaddress,amsmath,amsfonts,amssymb,aps]{revtex4}

\usepackage{graphicx}
\usepackage{dcolumn}
\usepackage{bm}
\usepackage[normalem]{ulem}
\usepackage{color}
\usepackage{amssymb}
\usepackage{amstext}

\begin{document}


\title{Localization Phenomena in Disordered Tantalum Films}

\author{Natalia Kovaleva}
\email{kovalevann@lebedev.ru}
\affiliation{Institute of Physics, Academy of Sciences of the Czech Republic,  Na Slovance 2, Prague 18221, Czech Republic}
\affiliation{Department of Physics, Loughborough University, Loughborough LE11 3TU, UK} 
\affiliation{Lebedev Physical Institute, Russian Academy of Sciences,  Leninskiy Prospekt 53, Moscow 119991, Russia}
\author{Dagmar Chvostova}
\affiliation{Institute of Physics, Academy of Sciences of the Czech Republic,  Na Slovance 2, Prague 18221, Czech Republic}
\author{A. Dejneka}
\affiliation{Institute of Physics, Academy of Sciences of the Czech Republic,  Na Slovance 2, Prague 18221, Czech Republic}

\date{\today}

\begin{abstract}
Using dc transport and wide-band spectroscopic ellipsometry techniques we study localization phenomena  in highly disordered metallic $\beta$-Ta films grown by rf sputtering deposition. The dc transport study implies non-metallic behavior (d$\rho$/d$T$\,$<$\,0), with negative temperature coefficient of resistivity (TCR). We found that as the absolute TCR value increased, specifying an elevated degree of disorder, the free charge carrier Drude response decreases, indicating the enhanced charge carrier localization. Moreover, we found that the pronounced changes occur at the extended spectral range, involving not only the Drude resonance, but also  the higher-energy Lorentz bands, in evidence of the attendant electronic correlations. We propose that the charge carrier localization, or~delocalization, is accompanied by the pronounced electronic band structure reconstruction  due to many-body effects, which may be the key feature for understanding the physics of highly disordered~metals.
\end{abstract}

\pacs{Valid PACS appear here}
\maketitle
\section{Introduction}

The phenomenon of negative temperature coefficient of resistivity (TCR) $\alpha_0$\,$<$\,0 in disordered metals \cite{Mooij}, found from temperature variation of their dc transport, $\rho(T)=\rho_0[1+\alpha_0(T-T_0)]$, was~considered and qualitatively evaluated by many researchers \cite{Jonson,Imry,Mott,Tsuei,Imry1,Gantmakher} in terms of weak localization effect \cite{Gorkov}. 
Generally, a weak localization interference effect takes place under the assumption that the electron wavelength is much smaller than the mean free electron path. At that point, the temperature is so low that the times of all inelastic processes $\tau_{\varphi}$, including electron-electron and/or electron-phonon scattering, during which the electron wave function coherence is preserved, are much longer than the elastic collision time $\tau_e$. In this case,  interference effects for the electron trajectories with self-intersection lead to the increased scattering probability, i.e., to the increased resistivity, which was estimated in the first approximation of perturbation theory \cite{Gorkov,Bergmann}. The diffusional character of electronic motion in a medium with impurities is generally satisfied at very low temperatures, in the regime of residual resistance. In principle, under the condition of increased disorder, scattering by static structural defects may occur more frequently than inelastic scattering by phonons in the appropriate temperature range. This~condition may be fulfilled in high-resistivity alloys, where the mean free path of conduction electrons in the process of elastic scattering at the static defects is of the order of magnitude of interatomic distance $l_e$ \textasciitilde{}~$a$ \textasciitilde~$k_{\rm F}^{-1}$ (where $k_{\rm F}$ \textasciitilde~$a^{-1}$ is the Fermi wavenumber). It was proposed that, due to this condition, weak localization correction may take place in disordered metals  and determine non-metallic character of their dc transport and negative TCR till room temperature \cite{Tsuei,Imry1,Gantmakher}. 

However, as a rule, the weak localization correction is small. 
Additionally, one-particle Anderson localization is expected in 
disordered metals under conditions of strong disorder. As a result, discrete localized electronic states appear near the Fermi level. One might 
expect the effect of fortifying electron correlations between the localized electrons. Thus,~the~interplay of disorder and electron correlations can lead to subtle many-body effects in localizing electrons, which represent fundamental challenges. These effects could be in evidence at optical frequencies. Here we propose an investigation of the dc transport of $\beta$-Ta films having large negative TCR~\cite{Read} in relation to their optical conductivity properties for various degrees of disorder. For that, we~suggest to use a spectroscopic ellipsometry approach, which was successfully applied in our earlier studies of electronic correlations in Mott-Hubbard insulators \cite{Kovaleva_lmo_prl,Kovaleva_lmo_prb,Kovaleva_yto_prb} and localization effects in the Kondo-lattice metal Ta$_2$PdSi$_3$ \cite{Kovaleva_TbPdSi}.

In the present study, $\beta$-Ta films, grown by rf sputtering deposition on a glass Sitall substrate, were investigated by four-point probe dc transport method and by spectroscopic ellipsometry with a~\mbox{J.A. Woollam} VUV-Gen II spectroscopic ellipsometer. The dc transport study  of the $\beta$-Ta films implies non-metallic behavior (${\rm d}\rho/{\rm d}T$\,$<$\,0) with large negative TCR values. In addition, our ellipsometry study demonstrates that the real part $\varepsilon_1(\omega)$ of the complex dielectric function of the studied $\beta$-Ta films shows the peculiar non-metallic behavior. We found that the observed non-metallic behavior is due to the presence of the pronounced Lorentz band around 2 eV, which is strongly superimposed with the free charge carrier Drude response at low photon energies. In addition, we found that  with increasing degree of disorder in the $\beta$-Ta films, as indicated by increasing absolute value of their TCR, the Drude contribution decreases, implying the enhanced charge carrier localization, and the intensity of the Lorentz band around 2 eV increases. Moreover, we found that the pronounced changes occur at the extended spectral range, involving also the higher-energy Lorentz bands. We suggest that this is indicative of the attendant electronic correlations in highly disordered metals due to many-body~effects.

The results of the present study can be useful in comprehension of many-body effects in mixed-valent magnetic compounds A$^{3+}_{1-x}$B$^{2+}_x$Mn$^{3+}_x$Mn$^{4+}_{1-x}$O$_3$ with A--B structural disorder, as well as in superconducting iron-based layered oxypnictides LaFeAsO$_{1-x}$F$_x$ with O--F structural disorder, and~other doped compounds of strongly correlated electron systems, on the basis of electronic localization due to structural disorder \cite{Varma,Ting,Boris}.

\section{Materials and Methods} \label{sec2}

Ta films were grown by rf sputtering deposition from 99.95\% pure Ta target on a glass substrate Sitall, commonly used for film substrates in microelectronics. In the process of  preparation for rf sputtering, the vacuum chamber was subjected to annealing at the temperature 200 $^\circ$C during one hour. The chamber base pressure before beginning the rf sputtering was about 2\,$\times$\,10$^{-6}$ Torr. In~the present rf sputtering experiments, a background Ar pressure was 6 $\times$ 10$^{-4}$ Torr, and the actual substrate temperature was about 80 $^\circ$C only. In our experiments we used the substrates with typical sizes of 15 $\times$ 5 $\times$ 0.6 mm$^3$. The film thickness was estimated by the deposition time, at which the material deposition rate was determined according to the procedure described in detail in Ref. \cite{Boltaev}. A set of the Ta films with different thickness was prepared. Usually, microstructure of the thin films grown by rf sputtering is thickness-dependent. 
Indeed, one would expect that  elevation of the film
surface temperature caused by rf discharge, as well as local
surface heating caused by atom bombardment with high
energy of about 1 eV, leads to the annealing effect and grain
growth along the film profile. Therefore, it is expected that the thinner Ta films, grown in the present study at the same rf sputtering conditions, will possess higher degree of disorder. By contrast, we suppose that concentration of oxygen defects, which can be caused
by oxygen presence at a background level of
10$^{-8}$--10$^{-9}$ mbar in using Ar gas, will be much less thickness
dependent, as its absorption occurs permanently in the
process of film growing. To protect from oxidation, the as deposited Ta films were capped  in situ with Al$_2$O$_3$ layer of 2.1 nm thick. The Al$_2$O$_3$ capping layer was prepared by rf sputtering of a single-crystal sapphire target. Our earlier studies \cite{Pud1,Pud2} show that even monolayers of active metals covered by the 2.1\,nm Al$_2$O$_3$ capping layer did not change their electronic properties during at least several years.

\begin{figure}[b]\vspace{-0.8em}
        \includegraphics[width=100mm]{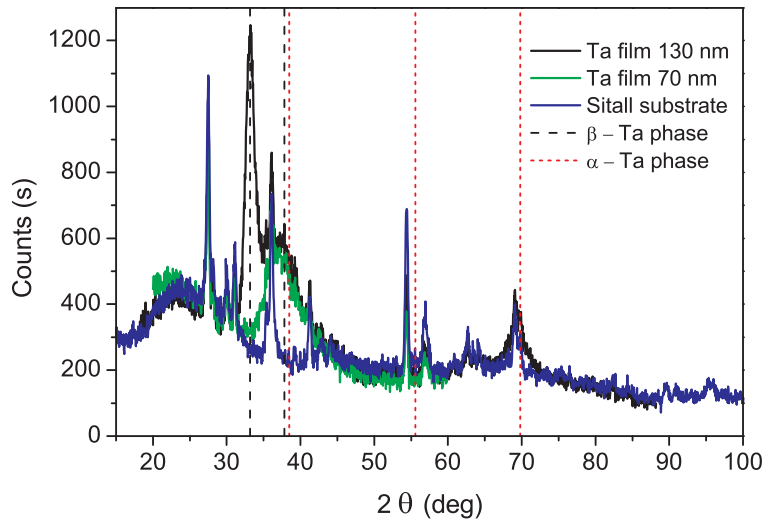}\vspace{-0.3em}
\caption{X-ray diffraction (XRD) analysis of the grown Ta/Sitall films. XRD spectrum of the blank Sitall substrate (blue line) and of the Ta films with thickness of 130 and 70 nm (black and green lines, respectively). The reflection positions corresponding to $\alpha$- and $\beta$-Ta phase are shown by the dashed~lines.}
\label{FigXRD}
\end{figure}
\vspace{-3pt}
A phase and a crystal structure of the sputtered Ta films were characterized by X-ray diffraction (XRD) analysis using Panalytical X'pert Pro MRD Extended Diffractometer (PANalytical, Almelo, the Netherlands) operating at 40 kV and 30 mA with Cu K$\alpha$ radiation. The $\theta$-$2\theta$ scans ranged between 15$^\circ$ and 100$^\circ$ with a step size of 0.02$^\circ$ and with time per step of 0.2 s were obtained. Figure \ref{FigXRD} shows the XRD spectrum of the blank Sitall substrate. The analysis of the XRD diffraction pattern showed that it is represented by the TiO$_2$ rutile phase. In Figure\,\ref{FigXRD} the XRD spectrum of the Ta film of 130 nm thick is superimposed with the XRD spectrum of the Sitall substrate. Here the peak with 2$\theta_1$ at 33.17$^\circ$  can be identified as (002) peak of $\beta$-Ta phase. The peak with 2$\theta_2$ at 37.8$^\circ$  may correspond to a sum of the peaks (202) and (212) of $\beta$-Ta, and thus showed  broadened appearance rather than a single distinct peak (see, for example, Figure\,2 of Ref. \cite{Shin}). 
The XRD study of the Ta film of 70 nm thick revealed the similar XRD pattern (see Figure\,\ref{FigXRD}), with notably decreased relative intensity of the (002) peak of $\beta$-Ta phase. From the present XRD study we may conclude that the grown Ta films are strongly disordered and rather represented by a mixture of the amorphous and fine-crystalline phases. At the same time, no clear traces of $\alpha$-Ta cubic phase were found in the film's XRD spectra. It is also important to note that the XRD analysis did not reveal any clear traces of Ta oxides in the grown Ta films. This signifies that the Ta oxides (such as TaO$_2$ and Ta$_2$O$_5$) can be available in the grown Ta films only at a relatively small concentration.

Sheet dc resistance of the rf sputtered Ta/Sitall films was measured in a wide temperature range on cooling from room temperature  down to 5\,K. For that, the film samples were mounted on a cold finger of the helium cryostat (RTI Cryomagnetic Systems, Moscow, Russia). The used multichannel electrical circuit allowed us to measure temperature dependence of sheet dc resistance simultaneously on several samples by four-point probe method.

Complex dielectric function spectra of the grown Ta/Sitall films were investigated in the wide photon energy range 0.8--8.5\,eV with a J.A. Woollam VUV-Gen II spectroscopic ellipsometer (J.A. Woollam Co., Lincoln, NE, USA). The~ellipsometry measurements were performed at two incident angles of 65$^\circ$ and 70$^\circ$ at room temperature.  Additionally, the ellipsometry measurements were performed on the blank Sitall substrate. With only a single angle of incidence, the raw experimental data are represented by real values of the ellipsometric angles $\Psi(\omega)$ and $\Delta(\omega)$. These values are defined through the complex Fresnel reflection coefficients for light-polarized parallel $r_p$ and perpendicular $r_s$ to the plane of incidence as follows
${\rm tan}\,\Psi\,e^{i\Delta}=\frac{r_p}{r_s}$.
The measured ellipsometric angles, $\Psi(\omega)$ and $\Delta(\omega)$, were~simulated using multilayer models available in the J.A. Woollam VASE software \cite{VASE}. 

\section{Results}
\vspace{-5pt}

\subsection{Temperature Study of dc Transport of the Ta Films}

\begin{figure}[b]\vspace{-0.8em}
        \includegraphics[width=100mm]{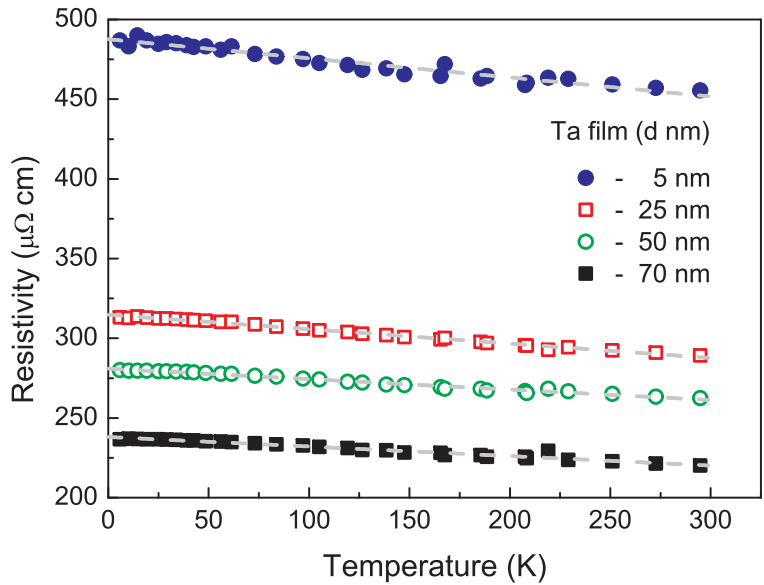}\vspace{-0.3em}
\caption{Temperature dependence of dc resistivity of the $\beta$-Ta films with thickness of 5.0, 25, 50, and 70 nm. Dashed lines represent the result of approximation by the linear dependence in the studied temperature range.}
\label{FigresSL}
\end{figure}

Figure\,\ref{FigresSL} shows temperature dependence of the dc resistivity of the Ta films with different thickness of 5.0, 25, 50, and 70 nm, estimated from nominal film thickness, actual sample geometry, and the distance between electrical contacts. One may notice that the dc resistivity (normalized to the dc resistivity at room temperature) increases by about 7\%--8\% with decreasing temperature from room temperature down to 5 K, demonstrating non-metallic character. In the temperature range 50\,K\,$\lesssim$\,T\,$\lesssim$\,180\,K, it is well approximated by the linear dependence, $\rho(T)$\,=\,$\rho_0[1+\alpha_0 (T-T_0)]$, where~$\alpha_0$\,$<$\,0 corresponds to the TCR, and $\rho_0$ is the resistivity value, obtained by the linear extrapolation to $T_0$\,=\,25\,$^\circ$C. The temperature dependence of the dc resistivity of the investigated Ta films shows small deviation from the linear approximation below 50\,K and above 180\,K. In the temperature range 50\,K\,$\lesssim$\,T\,$\lesssim$\,180\,K, for the Ta films with thickness 25, 50, and 70\,nm, we obtained the following values of $\alpha_0$ (TCR):  $-$345\,$\pm$\,9 ppm$\cdot$K$^{-1}$, $-$308\,$\pm$\,9 ppm$\cdot$K$^{-1}$, and $-$319\,$\pm$\,11 ppm$\cdot$K$^{-1}$ and $\rho_0$: 286\,$\pm$\,1\,$\mu \Omega$ $\cdot$cm, 259\,$\pm$\,1\,$\mu\Omega$$\cdot$cm, and 218\,$\pm$\,1\,$\mu \Omega\cdot$cm, correspondingly.
\vspace{-3pt}

The obtained data ($\alpha_0$,\,$\rho_0$) for the studied Ta films of different thickness well fit the range ($\alpha_0$,\,$\rho_0$) of the Mooij plot for disordered metals and alloys \cite{Mooij,Tsuei} and hold the similar trend. For example, the~increase in amplitude of the negative $\alpha_0$ value is accompanied by the increase in the resistivity $\rho_0$ with decreasing film thickness from 50 to 25\,nm. This notifies of the stronger degree of disorder in the thinner Ta film. However, the negative $\alpha_0$ values differ only very slightly for the Ta films with thickness of 50 and 70 nm. This is indicative of the nearly same degree of disorder in these films. Meanwhile, the $\rho_0$ value is notably less for the 70 nm Ta film. This could be due to the fact that the thicker Ta films are represented by a mixture of the amorphous and fine-crystalline phases, as we mentioned in Section \ref{sec2}. Besides, it is worth to note that the dc resistivity value in the studied Ta films is close to the critical value in disordered metals $\rho^*=\frac{\hbar k_{\rm F}}{ne^2l_e}=\frac{\hbar}{e^2}\frac{1}{k_{\rm F}}$\,$\simeq$\,300\,$\mu \Omega\cdot$cm, when an average mean free path of conducting electrons is approximately equal to the interatomic distance $l_e$ \textasciitilde{} $a$ \textasciitilde{} $k_{\rm F}^{-1}$. Interesting, in~accordance with our estimate, the dc resistivity value of the ultrathin Ta film of about 5.0 nm thick, $\rho_0$\,$\simeq$\,446\,$\pm$\,5\,$\mu \Omega\cdot$cm,  is about 1.5 times higher than the critical $\rho^*$ value (see Figure\,\ref{FigresSL}). A possible origin of this phenomenon will be further discussed.

\subsection{Spectroscopic Ellipsometry Study of the Ta films}

\begin{figure}[b]\vspace{-0.8em}
        \includegraphics[width=120mm]{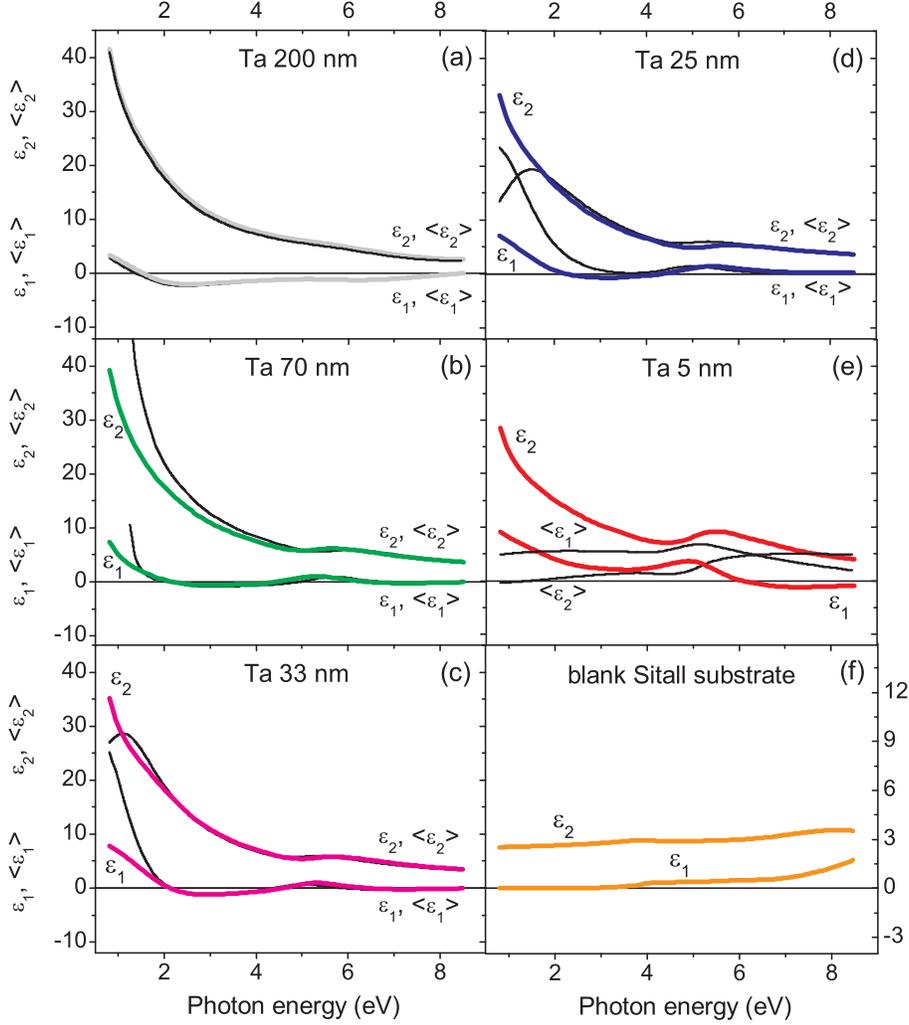}\vspace{-0.3em}
\caption{(\textbf{a}--\textbf{e}) The complex pseudo-dielectric function $\left\langle \varepsilon_1(\omega) \right\rangle$ and $\left\langle \varepsilon_2(\omega) \right\rangle$  (shown by solid black curves), obtained from the ellipsometry measurements (with an angle of incidence of 70$^\circ$) on the three layer system Al$_2$O$_3$(2.1 nm)/Ta/Sitall for the Ta films of different thickness of 200, 70, 33, 25, and~5~nm, respectively. Real $\varepsilon_1(\omega)$ and imaginary $\varepsilon_2(\omega)$ parts of the complex dielectric function of the Ta layer (shown by color curves), extracted from the Drude-Lorentz model (read details in the text); (\textbf{f})~The~complex dielectric function spectra of the blank Sitall substrate.}
\label{pseudoepsTa}
\end{figure}

Figure\,\ref{pseudoepsTa}a--e displays the complex pseudo-dielectric function $\left\langle \varepsilon_1(\omega) \right\rangle$ and $\left\langle \varepsilon_2(\omega) \right\rangle$ of the studied  three layer system Al$_2$O$_3$(2.1 nm)/Ta/Sitall for various thickness of the  $\beta$-Ta films of 200, 70, 33, 25, and 5\,nm,  obtained from the ellipsometry measurements. The complex pseudo-dielectric function $\left\langle \varepsilon_1(\omega) \right\rangle$ and $\left\langle \varepsilon_2(\omega) \right\rangle$ of the studied three layer system Al$_2$O$_3$(2.1 nm)/Ta/Sitall was calculated from the measured ellipsometric angles, $\Psi(\omega)$ and $\Delta(\omega)$, using the following expression  
\vspace{12pt}
\begin{eqnarray}
\left\langle \varepsilon \right\rangle=\left\langle \varepsilon_1 \right\rangle+i\left\langle \varepsilon_2 \right\rangle={\rm sin}^2\,\theta \left[ 1+{\rm tan}^2\,\theta \left( \frac{1-{\rm tan}\Psi e^{i\Delta}}{1+{\rm tan}\Psi e^{i\Delta}} \right)^2 \right],
\end{eqnarray}
where $\theta$ is an angle of incidence.

Dielectric function response of the investigated $\beta$-Ta films of different thickness was simulated using the three layer model Al$_2$O$_3$/Ta/Sitall. Complex dielectric function response, \mbox{$\tilde \varepsilon(\omega)=\varepsilon_1(\omega)+{\rm i}\varepsilon_2(\omega)$}, of a Ta layer was modelled by a Drude term, responsible for free charge carrier response, and a sum of contributions from Lorentz oscillators \cite{JWAllen} 
\begin{eqnarray}
\varepsilon=\epsilon_{\infty}-\frac{\hbar^2}{\varepsilon_0\rho\left( \tau E^2+{\rm i}\hbar E \right)}+\sum_j\frac{S_jE^2_j}{E_j^2-E^2-{\rm i}E\gamma_j}. \label{DrLor}
\end{eqnarray}

Here, $E\equiv\hbar\omega$, the Planck's constant $\hbar$, the vacuum dielectric constant $\varepsilon_0$, and the electron charge $e$ are
the physical constants, and $\epsilon_{\infty}$ is the core contribution to the dielectric function. The zero-frequency resistivity $\rho$\,=\,$\frac{m^*}{Ne^2\tau}$ and the mean scattering time $\tau$ were variable
parameters in fitting the Drude term, whereas $m^*$\,--\,the carrier effective mass and $N$---the charge carrier
concentration were adjustable non-fitted parameters. In addition, the $E_j$, $\gamma_j$, and $S_j$ were fitted parameters of the peak energy, full~width at half maximum, and oscillator strength of the $j^{th}$ Lorentz oscillator, correspondingly. 
\vspace{-3pt}

The measured ellipsometric angles, $\Psi(\omega)$ and $\Delta(\omega)$, were simulated using multilayer models available in the J.A. Woollam VASE software \cite{VASE}. In the simulation,  the dielectric function spectra obtained from our ellipsometry measurements on the blank Sitall substrate (shown in Figure \ref{pseudoepsTa}f) were substituted by the Gaussian functions. For the capping Al$_2$O$_3$ layer we used the complex dielectric function from Palik \cite{Palik}. The Ta layer thickness was fitted during the simulation. The~quality of the fit by the Drude-Lorentz model (Equation\,(\ref{DrLor})) was verified by the coincidence within the specified accuracy (less than 5\%) with the nominal Ta film thickness values and with the ellipsometric angles, $\Psi(\omega)$ and $\Delta(\omega)$, measured at two incident angles of 65$^\circ$ and 70$^\circ$. The resulting corresponding parameters for the $\beta$-Ta films of different thickness are summarized in Table \ref{Table1}. The dielectric function spectra, $\varepsilon_2(\omega)$ and $\varepsilon_1(\omega)$, of the studied $\beta$-Ta films with thickness of 200, 70, 33, 25, and 5\,nm, obtained from the modeling of the three layer system Al$_2$O$_3$/Ta/Sitall, are shown in Figure\,\ref{pseudoepsTa}a--e. One~may notice that the $\varepsilon_2(\omega)$ and $\varepsilon_1(\omega)$, resulting from the simulation, coincide with  the respective pseudo-dielectric function $\left\langle \varepsilon_1(\omega) \right\rangle$ and $\left\langle \varepsilon_2(\omega) \right\rangle$ for a bulk-like response only, as for the 200\,nm thick $\beta$-Ta film. In this case, the thick Ta film almost completely screens the substrate. One may also notice that the imaginary part $\varepsilon_2(\omega)$ of the complex dielectric function spectra decreases at low photon energies with decreasing the Ta film thickness. At the same time, the real part $\varepsilon_1(\omega)$ of the complex dielectric function of the studied $\beta$-Ta films demonstrates the peculiar non-metallic behavior, with up-turn to positive values.  

\begin{table}
\centering
\caption{The Drude--Lorentz parameters for the $\beta$-Ta films of different thickness, resulting from the modeling of the system Al$_2$O$_3$/Ta/Sitall by Equation\,(\ref{DrLor}). The values of $E$ and $\gamma$ are given in eV, $\rho^{opt}_0$ in $\mu \Omega$$\cdot$cm, $\tau$ in 10$^{-15}$ s, $m^*$ in units of free electron mass, and $N$ in 10$^{23}$ cm$^{-3}$.}
\begin{tabular}[c]{@{}c@{~~~~}c@{~~~~~}c@{~~}c@{~~}c@{~~}c@{~~}c@{~~}c@{~~}c@{~~}c@{~~}c@{~~}c@{}}
\hline
{\bf Contributions}&{\bf Parameters}&{{\bf 200\ nm}}~&{{\bf 70\ nm}}~&{{\bf 33\ nm}}~&{{\bf 25\ nm}}~&{{\bf 5\ nm}}\\\hline
Drude & $\rho_{\rm 0}^{opt}$ & 244 & 298 &307 &326 &386\\
& $\tau$ & 0.22 &0.23 &0.23 & 0.24 & 0.17\\ 
& $N$ & 0.66 & 0.54&0.51 & 0.45 & 0.54\\
& $m^*$ & 1 &1 &1 & 1 & 1\\
& $\epsilon_{\infty}$ & 0.97& 1.23& 1.14 & 1.17 & 1.60\\ \hline
& $E$  & 2.0 (3) &1.9 (9)&2.1 (6) & 2.1 (7) & 2.2 (3) \\
Lorentz oscillator & $S_{\rm osc}$ & 9.9 (8) &14.0 (0)& 13.0 (0) & 12.4 (4) & 8.3 (3)\\  
& $\gamma$ & 3.1 (6) &4.6 (5)&3.2 (2) & 3.5 (1) & 3.7 (9)\\ \hline
& $E$  & 4.1 (5)  &3.4 (4)&3.7 (7)  & 3.5 (9) & 3.3 (7)\\
Lorentz oscillator & $S_{\rm osc}$ & 2.2 (4)   &1.4 (7)& 0.9  (3)  & 1.0 (2) & 1.9 (6)\\
& $\gamma$ & 4.1 (8)  &3.6 (7)& 2.5 (3) & 2.5 (5) & 3.1 (6)\\ \hline
& $E$  & 5.9 (3) &5.9 (2)&5.8 (9)   & 6.0 (3) & 5.5 (5)\\
Lorentz oscillator & $S_{\rm osc}$ & 1.1 (0)&0.4 (0)& 0.5 (3)  & 0.5 (1) & 1.1 (7)\\
& $\gamma$ & 3.4  (0) &1.5(3)& 1.7 (2) & 1.8(0) & 1.5 (6)\\ \hline
& $\omega$  & -- &7.3 (1)&7.3 (4)  & 7.6 (1) & 6.6 (9) \\
Lorentz oscillator & $S_{\rm osc}$ & -- &0.9 (9)& 1.1(3) & 1.1(5) & 1.1(8)  \\
&$\gamma$ & -- &3.8 (1)& 3.9 (6) & 4.0 (6) & 2.5 (8) \\ \hline
\end{tabular}
\label{Table1}       
\end{table}

\begin{figure}[b]\vspace{-0.8em}
        \includegraphics[width=120mm]{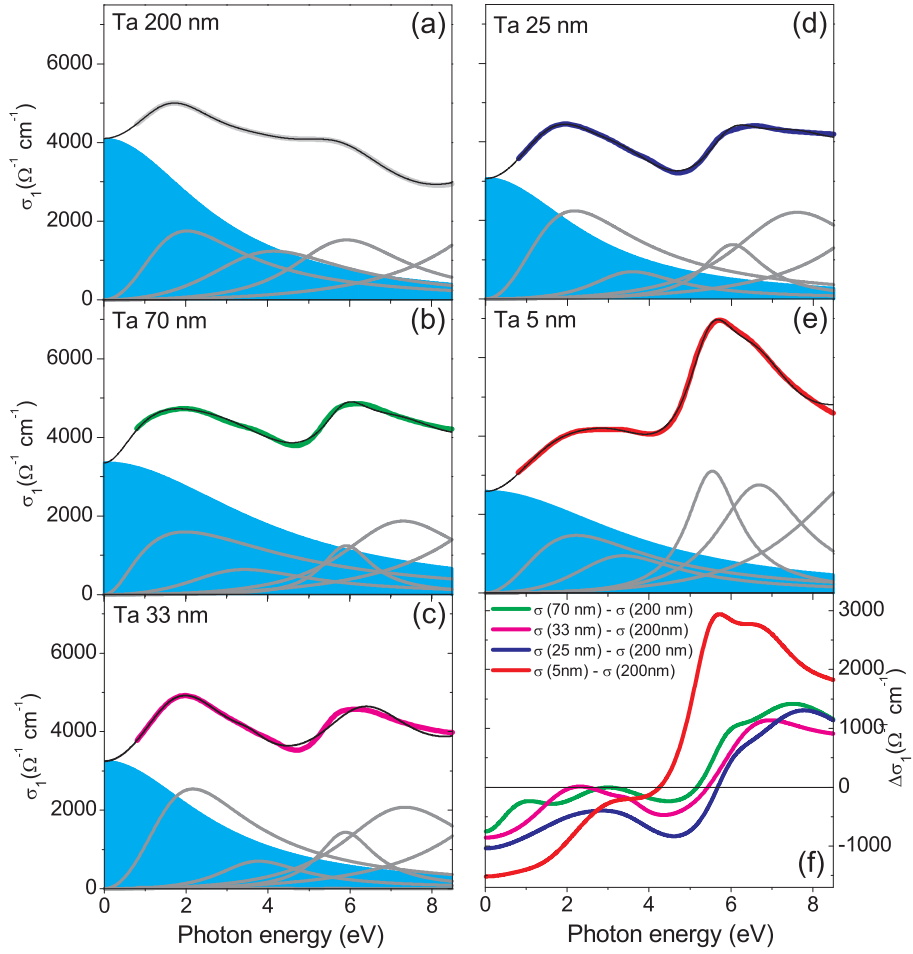}\vspace{-0.3em}
\caption{(\textbf{a}--\textbf{e}) Optical conductivity, $\sigma_1(\omega)=\frac{1}{4\pi}\omega \varepsilon_2(\omega)$, of the studied Ta films of different thickness (shown by color curves). The contributions from the Drude term (cyan shaded area) and Lorentz oscillators (solid gray lines) are explicitly displayed. Solid black lines show their summary contribution to the optical conductivity; (\textbf{f}) Difference for the optical conductivity spectra of the single layer Ta films with decreasing film thickness.}
\label{epsTa}
\end{figure}

Figure\,\ref{epsTa}a--e shows the associated optical conductivity spectra, $\sigma_1(\omega)=\frac{1}{4\pi}\omega \varepsilon_2(\omega)$, where contributions from the Drude term and Lorentz oscillators (calculated from the corresponding parameters listed in Table \ref{Table1}) are explicitly displayed. One can see that the dielectric function response of the studied Ta films at low photon energies is represented by the Drude resonance and the intense Lorentz band peaking around 2 eV, which are strongly superimposed (see Figure\,\ref{epsTa}a--e). We found that, in the case of their strong superposition, the non-metallic behavior of the real part $\varepsilon_1(\omega)$ of the dielectric function is dictated by  the Kramers-Kronig consistency of the applied Drude-Lorentz model (Equation\,(\ref{DrLor})). In addition, in the range of interband transitions, we identified optical transitions, where the dominant contribution can be associated with the Lorentz bands peaking around 4 and 6\,eV (see Table \ref{Table1}). From the detailed dispersion analysis, we discovered that the Drude contribution decreases with decreasing the Ta film thickness from 200 to 25\,nm (see Figure\,\ref{epsTa}a--d,f), which indicates on the enhanced charge carrier localization. Moreover, we found that pronounced changes occur at the extended spectral range, involving not only the Drude resonance but also the dominant Lorentz bands with the energies $E_j$\,$\sim$\,2, 4, and 6--8\,eV. Here, the decrease of the Drude contribution with decreasing the Ta film thickness from 200 nm is accompanied by the increase of oscillator strength of the Lorentz band at 2\,eV and by the decrease of oscillator strength of the Lorentz band at 4\,eV (see Table \ref{Table1}). In~addition, one can see that oscillator strength of the Lorentz bands at around 6--8\,eV increases with decreasing the Ta film thickness (see Figure\,\ref{epsTa}a--d). And, for the thinnest investigated Ta film of about 5.0 nm thick, we noticed a sharp decrease of the Drude contribution (see Figure\,\ref{epsTa}e). In line with the dc transport study, a reciprocal value of the Drude dc limit, $\rho_0^{opt}$\,$\simeq$\,390 $\Omega$$\cdot$cm$^{-1}$, is well above the critical $\rho^*$ value for disordered metals. Interesting that simultaneously the intensity of the optical band with the energy $E_j$\,$\sim$\,2 eV becomes essentially suppressed. Instead, one can follow the pronounced increase of the higher-energy Lorentz band at $E_j$\,$\sim$\,5.5 eV.

\section{Discussion}

Using the obtained results, we discuss the dc electron transport properties of the studied $\beta$-Ta films of different thickness. A valuable information can be obtained from the Drude term parameters listed in Table \ref{Table1}. It is generally known that mean path of conduction electrons $l$ is related to mean scattering time $\tau$ as $l=\frac{\tau\hbar(3\pi^2N)^{1/3}}{m^*}$, where $N$ is the concentration of free charge carriers, and $m^*$ is the effective electron mass. Taking the Drude parameters $\tau$\,$\simeq$\,$0.23~\times$ 10$^{-15}$\,s and $N$\,$\simeq$\,$0.54~\times$ 10$^{23}$\,cm$^{-3}$ from Table \ref{Table1}, we determine the average mean free path $l$(70\,nm)\,$\simeq$ 3.12\,\AA\, for the Ta film of 70\,nm thick. The estimated mean free path is smaller than the in-plane tetragonal ($P42/mnm$) $\beta$-Ta lattice parameter a\,=\,5.34\,\AA\,\cite{Read}. It can be rather compared with the cubic ($Im3m$) $\alpha$-Ta lattice parameter 3.31--3.33\,\AA\,\cite{Read}. The Drude parameters listed in Table \ref{Table1} suggest that the average mean free electron path slightly increases with increasing the Ta film thickness, thus $l$(200\,nm)\,$\simeq$\,3.19\,\AA,\, and decreases with decreasing the Ta film thickness, for example, $l$(25\,nm)\,$\simeq$\,3.06\,\AA. This means that the studied $\beta$-Ta films are strongly disordered, almost amorphous, where the high scattering probability restricts the average mean free electron path to $l$\,$\simeq$\,3.06--3.19 \AA. At the same time, using the Drude parameters, one can estimate the dc conductivity, $\sigma=\frac{Ne^2\tau}{m^*}$. For example, this gives $\sigma$(200\,nm)\,$\simeq$\,4100\,$\Omega^{-1}\cdot$\,cm$^{-1}$, $\sigma$(70\,nm)\,$\simeq$\,3500 $\Omega^{-1}\cdot$\,cm$^{-1}$, and $\sigma$(25\,nm)\,$\simeq$\,3000\,$\Omega^{-1}\cdot$\,cm$^{-1}$. Their reciprocal values correspond to the optical resistivity data $\rho^{opt}_0$ given in Table \ref{Table1}.  

On the other hand, we found that the temperature variation of the dc transport, \mbox{$\rho(T)=\rho_0\left[ 1+\alpha_0(T-T_0)\right]$}, of the studied $\beta$-Ta films shows non-metallic behavior (${\rm d}\rho/{\rm d}T$\,$<$\,0) with negative TCR ($\alpha_0$). The determined  ($\alpha_0$,\,$\rho_0$) values, as well as ($\alpha_0$,\,$\rho^{opt}_0$) values, well fit the range of the Mooij plot for highly disordered or amorphous metals having negative TCR and show the similar trend, implying that the physics of the studied Ta films is driven by static disorder \cite{Mooij,Tsuei}. Moreover, the~acquired ($\alpha_0$,\,$\rho^{opt}_0$) dependence indicates on the general trend of a stronger degree of disorder in the thinner Ta films.

\begin{figure}[b]\vspace{-0.8em}
        \includegraphics[width=110mm]{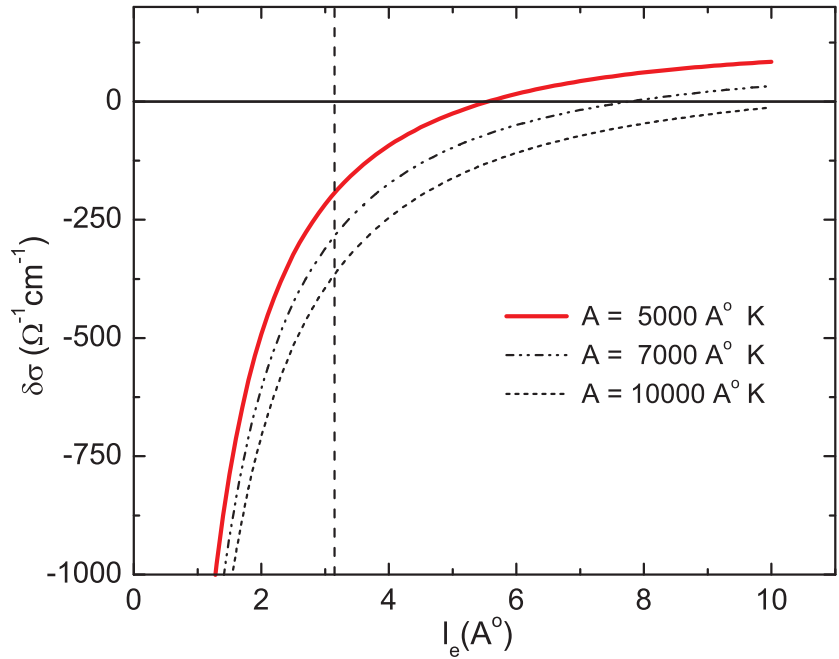}\vspace{-0.3em}
\caption{Weak localization correction $\delta\sigma$(3D) to the Boltzmann conductivity (in accordance with Equation\,(\ref{WLC})) as a function of the mean free electron path between elastic collisions $l_e$. For the estimate at $T$\,=\,300\,K, the mean free electron path between inelastic collisions $l_\varphi$\,=\,$AT^{-P}$,\,$P$\,=\,1 \cite{Tsuei}. The estimate is given for different values of the parameter $A$.}
\label{FigWLC}     
\end{figure}

The correlation between TCR ($\alpha_0$) and resistivity $\rho_0$, and, in particular, the phenomenon of negative TCR in disordered metals \cite{Mooij}, were considered and qualitatively evaluated by many researchers \cite{Jonson,Imry,Mott,Tsuei,Imry1,Gantmakher} in terms of weak localization effect \cite{Gorkov}. As a rule, the weak localization correction  is small and in 3D metals is determined as follows
\begin{eqnarray}
\delta \sigma(T)=\frac{e^2}{\pi^2\hbar}\left[ \frac{1}{L_\varphi(T)}-\frac{1}{l_e}\right],
\label{WLC}
\end{eqnarray}
where $l_e$ and $l_\varphi$ are an average mean free path of electrons between elastic and inelastic scattering, respectively, and $L_\varphi=\left( \frac{1}{3}l_\varphi l_e \right)^{1/2}$ is the diffusion length \cite{Tsuei}. For an estimate at $T$\,=\,300\,K one can admit that $l_\varphi(T)=AT^{-P}$, $P$\,=\,1. We take for $A$\,=\,5000 \AA\,K from the results by Tsuei \cite{Tsuei}. Then, $l_\varphi(300\,{\rm K})$\,$\simeq$\,17\,\AA, and for the mean free path $l_e$\,$\simeq$\,3--5\,\AA\,the diffusion length will be $L_\varphi(300\,{\rm K})$\,$\simeq$\,4.1--5.3\,\AA. By using Equation\,(\ref{WLC}), we can estimate $\delta \sigma(l_e,300\, {\rm K})$ in disordered metals, using $l_\varphi(T)=AT^{-P}$, $P$\,=\,1 \cite{Tsuei}. In~Figure\,\ref{FigWLC} we plot $\delta \sigma(l_e,300\, {\rm K})$ as a function of the mean elastic free electron path $l_e$ for different values of the parameter $A$\,=\,5000, 7000, and 10,000\,\AA\,K. As one can see from the plot, for the average mean free electron path $l$\,$\simeq$\,3.06--3.19\,\AA, peculiar of the studied disordered $\beta$-Ta films, the effect of weak localization correction could achieve $\delta\sigma$\,$\sim$\,(180--360)\,$\Omega^{-1}\cdot$\,cm$^{-1}$.
Thus, on one hand, the acquired in the present study ($\alpha_0$,\,$\rho_0$) values, as well as ($\alpha_0$,\,$\rho^{opt}_0$) values, well fit the range of the~Mooij plot for highly disordered or amorphous metals having negative TCR. On the other hand, we~evaluated that the effect of weak localization correction could achieve here $\delta\sigma$\,$\sim$\,(180--360)\,$\Omega^{-1}\cdot$\,cm$^{-1}$ only. However, this is far less than the estimated above dc conductivity variation in the studied Ta films of different thickness, namely, $\sigma$(200\,nm)\,$\simeq$\,4100\,$\Omega^{-1}\cdot$\,cm$^{-1}$, $\sigma$(70\,nm)\,$\simeq$\,3500\,$\Omega^{-1}$\,$\cdot$\,cm$^{-1}$, and~$\sigma$(25\,nm)\,$\simeq$\,3000\,$\Omega^{-1}$\,$\cdot$\,cm$^{-1}$. Therefore, there must be some additional reasons, which cause  the observed pronounced dc conductivity changes.

One simple reason may be related to the effect of oxygen contamination and possible formation of chemically stable Ta oxides, such as TaO$_2$ and/or Ta$_2$O$_5$ in the grown Ta films. However, the~XRD analysis of the grown Ta films did not reveal any clear traces of Ta oxides in the grown Ta films.
In~addition, as we mentioned, we did not expect that concentration of oxygen defects, which can be caused by oxygen presence at a background level in using Ar gas, will be notably thickness dependent. Moreover, the bandgap value for sub-stoichiometric and stoichiometric TaO$_2$ and Ta$_2$O$_5$ films is reported to be comprised in the spectral range 3.9--5.5\,eV \cite{Woollam,Ta2O5,TaO2}. In the present study, we observed a~pronounced increase of the higher-energy Lorentz band around 5.5 eV, simultaneously with a sharp decrease of the Drude contribution only for the thinnest Ta film of about 5.0 nm thick (see Figure\,\ref{epsTa}e). In fact, from the present spectroscopic ellipsometry study, we conclude that the non-metallic character of the dielectric function response of the grown $\beta$-Ta films is associated with the presence of the intense Lorentz band peaking around 2 eV, which is strongly superimposed with the free charge carrier Drude response at low photon energies (see Figures\,\ref{pseudoepsTa}a--e and \ref{epsTa}a--f). Moreover, we found that  with increasing degree of disorder in the thinner Ta films, as indicated by increasing absolute value of their TCR, the intensity of the Lorentz band at 2 eV increases, whereas the intensity of the Lorentz band at 4 eV decreases (see Table \ref{Table1}). The observed decrease  of the Lorenz band at 4 eV in the thinner Ta films indicates that the effect of oxygen contamination is comparatively small and therefore cannot be responsible for the observed optical conductivity changes. 

However, we noticed that our data provide apparent analogies to the dielectric function behavior reported for TaN$_x$ layers grown at different deposition temperatures (200--400\,$^\circ$C) \cite{Mistrik}. 
By the way, here, the content of oxygen in the films was estimated at 3\% only. The growth phase map \cite{Shin} suggests that, for the deposition conditions, the film composition is single-phase metastable $\delta$-TaN$_x$ for all deposition temperatures from 100 to 600\,$^\circ$C (see Figure 7 of Ref.\,\cite{Shin}).
The reported TaN$_x$ dielectric function range from metallic to non-metallic character, depending on the substrate temperature. The~most pronounced metallic character exhibits the TaN$_x$ layer deposited with the highest
substrate temperature of 400\,$^\circ$C. In fact, this points on the defect annealing effect, indicating that the observed behavior of the dielectric function of  the TaN$_x$ layers studied by Mistrik  et al. \cite{Mistrik} is driven by static disorder. In addition, the optical properties of TaN were studied by spectroscopic ellipsometry and detailed DFT ab initio band structure calculations by Matenoglou et al. \cite{Matenoglou}. It was shown that, along
with the contribution from the Drude term describing the intraband absorption within the Ta 5d $t_{2g}$ band that intercepts the Fermi level, the additional Lorentz band was also observed at low photon energies around 1.9 eV.

Thus, we found that  with increasing degree of disorder in the thinner Ta films, as indicated by increasing absolute value of their TCR, the Drude contribution due to intraband absorption within the Ta 5d $t_{2g}$ band at the Fermi level decreases due to localization effects. The associated optical spectral weight is recovered in the range of the higher-energy Lorentz bands at 2, 4, and 6--8 eV. Recently, we~have shown that electron localization in the formed metallic magnetic clusters in the Kondo-lattice metal Tb$_2$PdSi$_3$ leads to opening the pseudo-gap in the conduction band of itinerant electrons and appearance of the Mott-Hubbard-like interband optical transitions \cite{Kovaleva_TbPdSi,Kovaleva_lmo_prl,Kovaleva_lmo_prb,Kovaleva_yto_prb}. In the disordered Ta films, the observed low-energy transitions at around 2 and 4 eV may have some association with d$^3$d$^3$ $\rightleftarrows$ d$^2$d$^4$ Mott-Hubbard-like electron correlations \cite{Kovaleva_lmo_prl,Kovaleva_lmo_prb}. In this case, the observed decrease of the free charge carrier conductivity (Drude) with increasing the disorder degree, accompanied by the changes of intensities of the Mott-Hubbard-like bands, may indicate opening the pseudo-gap in the conduction band of itinerant electrons and global reconstruction of the band structure in disordered metals, associated with the many-body Anderson localization effects. This leads to the accelerated decrease of the Drude conductivity with increasing degree of disorder and can explain the observed pronounced dc conductivity variation in the studied Ta films. Though, this hypothesis requires further verification by theoretical band structure calculations for $\beta$ Ta, as well as comprehensive experimental investigation of the temperature dependence of dielectric function spectra, by using, for example, spectroscopic ellipsometry approach.

Interestingly, for the ultrathin Ta film of about 5.0 nm thick, the Drude dc conductivity limit drops below the weak localization limit for disordered metals $1/\rho_0^*$\,$\simeq$\,3300\,$\Omega^{-1}$\,$\cdot$\,cm$^{-1}$. Simultaneously, the~global band structure reconstruction occurs (see Figure\,\ref{epsTa}e,f). Namely, the intensity of the optical band with the energy $E_j$\,$\sim$\,2 eV becomes essentially suppressed, instead, the higher-energy Lorentz band at $E_j$\,$\sim$\,5.5 eV becomes tremendously pronounced. One simple reason for this may be related to the effect of oxygen contamination and the formation of chemically stable Ta oxides, such as TaO$_2$ and/or Ta$_2$O$_5$, which, in principle, may be peculiar for the ultrathin Ta film, as we discussed earlier. However, another possible origin, associated with the nearly many-body localized state, cannot be excluded. According~to the results of our atomic force microscopy study of the Sitall substrate, its profile shows the height variation of 1--3 nm, (which constitutes from 20\% to 60\% of the nominal film thickness) at the lateral scale 50--100 nm \cite{Stupakov}. We may 
propose that the roughness profile, with the peculiar long-range 
disorder, compared to the Ta lattice spacing, may enhance the Anderson localization, resulting in the formation of the nearly many-body localized electron state \cite{Serbyn}. In this case, the system may exhibit spectral gaps having different hierarcy of scales. The issue of the surface roughness on the possible formation of the many-body localized electron state requires more experimental and theoretical~investigations.  
\section{Conclusions}

In summary, here we have studied dc transport and complex dielectric function spectra of highly disordered metallic $\beta$-Ta films. Temperature variation of the dc transport of the studied $\beta$-Ta films shows non-metallic behavior, $\rho(T)=\rho_0\left[ 1+\alpha_0(T-T_0)\right]$, with negative TCR ($\alpha_0$). The determined ($\alpha_0$,\,$\rho_0$) values well fit the range of the Mooij plot for highly disordered or amorphous metals and show the similar trend, indicating that the physics of the studied $\beta$-Ta films is driven by static disorder \cite{Mooij,Tsuei}. We found that with
increasing the TCR absolute value, specifying elevated degree of disorder, the free charge carrier Drude response decreases, implying the enhanced charge carrier localization. Moreover, we found that the pronounced changes occur at the extended spectral range, involving not only the Drude resonance, but also  the higher-energy Lorentz bands, in evidence of the attendant electronic correlations. We propose that the charge carrier localization, or delocalization, is accompanied by the pronounced electronic band structure reconstruction due to many-body effects, which may be the key feature for the physics of highly disordered metals.\\

We thank Anton Bagdinov for participation in the dc transport measurements, Viktor Martovitsky for the XRD measurements and analysis, and Fedor Pudonin for the Ta films growing. 
This work was supported by grant 15-13778S of the Czech Science Foundation.

\end{document}